\documentstyle[preprint,aps,epsfig,prb]{revtex}
\tightenlines
\begin{document}
\title {\bf Tensile Strength of Carbon Nanotubes under Realistic Temperature and Strain Rate}
\author{Chenyu Wei $^{a,b}$, Kyeongjae Cho$^a$, and Deepak
Srivastava $^b$}
\address{Department of Mechanical Engineering, Stanford University,
California 94305, $^a$ \\ NASA Ames Research Center, MS T27A-1,
Moffett Field, California 94035 $^b$}

\date{\today}

\maketitle

\begin{abstract} {Strain rate and temperature dependence of the tensile strength of single-wall carbon nanotubes has been investigated with molecular dynamics simulations. The tensile failure or yield strain is found to be strongly dependent on the temperature and strain rate. A transition state theory based predictive model is developed for the tensile failure of nanotubes. Based on the parameters fitted from high-strain rate and temperature dependent molecular dynamics simulations, the model predicts that a defect free $\mu m$ long single-wall nanotube at 300K, stretched with a strain rate of $1\%/hour$, fails at about $9 \pm 1\%$ tensile strain. This is in good agreement with recent experimental findings.}
\end{abstract}

Carbon nanotubes (CNTs) have very high strength and stiffness under axial stresses, but are very flexible under non-axial stresses. The theoretical 
and experimental values of the Young's modulus are found to be around 1 TPa. 
For large tensile stresses, the atomistic molecular dynamics (MD) simulations performed at very high strain rates (limited by time-scale of the phenomenon that can be simulated in MD) show tensile failure strains to be as high as 30$\%$.\cite{yakobson97} Whereas, recent experiments report much smaller tensile failure strains in a variety of scenarios. Walters et al \cite{smalley99} have reported a failure strain of $5.8\%$ on single walled CNT (SWCNT) ropes. Yu et al \cite{yu001} have reported a similar value of maximum $5.3\%$ strain for failure at room temperature, and have measured strain for failure of multi-walled CNTs (MWCNT)\cite{yu002} to be as high as $12\%$.

Earlier theoretical and numerical simulation studies have shown that, under large tensile strains, Stone-Wales(SW) bond rotations result in the formation of pentagon-heptagon pair (5775) defects on the nanotubes (Figure 1). The formation energy of such defects decreases with the increasing tensile strains and these defects are energetically favorable at tensile strains larger than $5\%$,\cite{nardelli98} However, the large energy barriers (of about 8-9 eV) to the formation of SW bond rotations in static calculations\cite{crespi98} do not readily explain the experimentally measured low tensile strength of SWCNT ropes or MWCNTs. The high strain rates used in the MD simulations, on the other hand, are generally of the order of $pico$ $second^{-1}$ or faster, which are unrealistic, as compared to the experimental strain rates of $minute^{-1}$ or slower. A clear deficiency thus exists in a direct comparison of MD simulated tensile strength with what is measured in experiments. It is not clear if the differences are due to the real physical and chemical mechanistic reasons, or due to the limitations of the extremely fast strain rates used in the atomistic MD simulations. 

A typical time step in MD simulations can not exceed a fraction of the vibrational periods of atoms or molecules under investigations, and is 
generally of the order of few femto-seconds or less. Hundreds of millions of time steps, therefore, are needed to simulate typical nanosecond dynamic processes in atomic systems. For thousands of atoms, interacting with complex many-body force field functions, the MD simulation is computationally intensive and is not always accessible for routine investigations. One way to accelerate the kinetic processes, that occur during MD simulations, is to increase the temperature. The transition time, for an activated process, to go from one 
state to another is given by the Arrhenious relation $t = {1 \over \nu} e^{E_{\nu}/k_{B}T}$, where $E_{\nu}$ is the activation energy and $\nu$ is the effective vibration or attempt frequency of the transition. At high temperatures the larger kinetic energy increases the rate of the process to overcome the barriers and the transition time is thus shortened. If the tensile yielding of a material can be described by a series of activated processes of defect formations. It may be possible to describe the tensile yielding or failure of CNTs with an Arrhenious type model incorporating both short time and high temperature defect formation behavior, as well as long time and low temperature behavior. We have used temperature and strain rate dependent (with in MD time-scale) MD simulations to propose and parameterize a predictive model for tensile yield strain of micron long SWCNTs at experimentally feasible strain rates and temperatures. 

The MD simulations involve continuous straining of nanotubes by moving the end atoms under given strain conditions, such as under tensile and compression strain, and letting the nanotube relax during the dynamics at constant temperatures and strain rates. Tersoff-Brenner potential \cite{tersoff88,brenner90} is used to model Carbon-Carbon (C-C) interactions. A 60$\rm \AA$ long (10,0) nanotube with 600 atoms is used for the simulation studies, and the CNT is strained at rates varying from $10^{-6}ps^{-1}$ to $10^{-2}ps^{-1}$ and at temperatures ranging from 300K to 2400K.

In Figure 2, the strain energy per atom for different temperatures and strain  rates, as a function of tensile strain, is plotted. The abrupt deviation from the elastic behavior, at high strain values (and we define such strains as yield strains or failure strains), is due to the formation of SW 5775 defects followed by necking, thinning and eventual breaking of the nanotubes. For strain values higher than the yield strain, the perfect hexagonal configurations of the carbon atoms in the CNTs are destroyed. This is in qualitative agreement with the previous MD simulations.\cite{yakobson97,yakobson98} Figure 2 also shows that the yield strain (and the corresponding tensile strength) decreases at higher temperatures and at slower strain rates. Initial MD studies by Yakobson et. al. \cite{yakobson97} showed similar behaviors but with very high yield strains due to the very high strain rates used in their simulations. Our simulations with a larger variation of strain rates and temperatures show the changes in the yield strain (or tensile strength) in a wider range. For example, at slow strain rate ($3 \times 10^{-5} /ps$) and high temperature of 2400K, the nanotubes yield at about $6\%$ in our simulations. The yield strain as a function of strain rate, at different temperatures, is shown in Fig. 3. The linear dependence at each temperature indicates that the Arrhenious behavior is valid and the tensile yielding of nanotubes can be described by a theory of activated processes, i.e., the transition of the system from the initial pre-yielding state to the final post-yielding state occurs through a series of activated processes with an effective barrier defining the characteristics of the whole phenomenon.

According to the Eyring's theory the Arrhenius formula for the strained system can be modified as \cite{eyring}

\begin{equation} t = {1\over
n_{site}}{1\over{\nu}}e^{{E_{\nu}-VK\epsilon}\over k_{B}T} ,
\end{equation}

where $\epsilon$ is local strain, $n_{site}$ is number of sites available for state transition, $K$ is force constant and $V$ is the activation volume. Thus the strain rate required for a transition at a certain strain is:

\begin{equation}
 {1 \over \dot{\epsilon}} = {1\over n_{site}}{1\over \dot\epsilon_{0}}e^{{E_{\nu}-VK \epsilon}\over k_{B}T} ,
\end{equation}

where $\dot{\epsilon}$ is strain rate and $\dot\epsilon_{0}$ is a constant related with vibration or attempt frequency. The yield strain as a function of temperature and strain rate, thus, can be expressed by inverting Eq. (2) as,

\begin{equation}
\epsilon_Y = {E_{\nu}\over{VK}} + {k_{B}T\over{VK}}\ln{\dot{\epsilon}\over n_{site}\dot\epsilon_{0}} .
\end{equation}

For nanotubes under tensile strain, we have observed that the yielding process induces a sequence of S-W type bond rotations within the connected local regions. These multiple transitions lead to a collective kinetic activation mechanism of the failure of CNTs strained beyond elastic limit. The combined rate for $N$ multiple transitions thus can be of the form $1\over{N t}$, where $1\over t$ is the rate for a single transition and for simplicity we have assumed that the rate for each successive transition can be expressed by an averaged value of $1/t$. An averaged effective activation energy ${\overline E}_{\nu}$ for the multiprocess thus replaces $E_{\nu}$ for the single process, and Eq. (3), can be modified as:

\begin{equation}
\epsilon_Y = {{\overline E}_{\nu}\over{VK}} + {k_{B}T\over{VK}}\ln{N \dot{\epsilon}\over n_{site}\dot\epsilon_0} . 
\end{equation}

To fit the data from the simulations, Eq. (4) can be rewritten as:
\begin{eqnarray}
\epsilon_Y & = & ({{\overline E}_{\nu}\over{VK}}-{k_{B}T\over{VK}}\ln({\dot\epsilon_0 {n_{site}}\over N \Delta\epsilon_{step}})) + {k_{B}T\over{VK}}\ln{\dot{\epsilon}\over\Delta\epsilon_{step}} \nonumber \\
& = & A(T) + B(T)\ln{\dot{\epsilon}\over\Delta\epsilon_{step}} ,
\end{eqnarray} 

where $A(T)$ and $B(T)$ are defined as $A(T)= {{\overline E}_{\nu}\over{VK}}-{k_{B}T\over{VK}}\ln({{n_{site}}\dot\epsilon_0 \over N \Delta\epsilon_{step}})$ and $B(T)={k_{B}T\over{VK}}$. $\Delta\epsilon_{step}$ is the change of strain at each step used in simulations. In Fig. 4, the functional dependence of $A(T)$ and $B(T)$ is obtained over the entire temperature range by fitting the linear dependence to the simulation data. The linear dependence of $B(T)$ on the temperature indicates that the transition state theory (within Eyring's model) is valid for the yielding of the strained CNTs. From the dependence of $B(T)$ 
on $T$, we get $VK$ = 18.04 eV. For one-dimensional CNTs, with Young's modulus giving the force constant to be 1 TPa, the activation volume comes out to be 2.88${\rm \AA}^3$. This corresponds to a typical atomic volume of a carbon atom within a nanotube. The fitting of the coefficients in $A(T)$ gives ${{\overline E}_{\nu}\over{VK}} = 0.20$ and ${k_{B}\over VK}\ln({{n_{site}}\dot\epsilon_0 \over N \Delta\epsilon_{step}}) = 3.66 \times 10^{-5}$. These values correspond to the average kinetic activation energy ${\overline E}_{\nu}$ of 3.6 eV for the successive multiple S-W bond rotations. In our simulations, with $n_{site} \approx$ $ 600$ and $\Delta\epsilon_{step} = 0.0025$ or $0.25\%$, $\dot\epsilon_0 \over N$ comes out to be about 8 $\times 10^{-3}$ $ ps^{-1}$.  $\epsilon_Y$ is not dependent on $\Delta\epsilon_{step}$, as the effects cancel out in $A(T)$ and $B(T)$.  There is always a possibility for the inverse transitions which have rates scaled as $e^{-{2 VK \epsilon \over k_{B}T}}\dot{\epsilon}$, and are omitted as ${VK \epsilon \over k_{B}T}>> 1$ in our study.

The yield strain of the CNTs, under tensile stress with experimentally feasible strain rates, can be estimated or predicted from the above model. 
For example, at T = 300K and at a strain rate of about $1\%$ $per$ $hour$ the yield or failure strain for a 6nm long (10,0) CNT comes out to be about $11\pm1\%$ as according to $\epsilon_Y (\%)= 20.21 + (0.21 \pm 0.02)\ln{\dot{\epsilon}}$ $(ps^{-1})$ from the data shown shown in Fig. 3. Furthermore, the effect of experimentally realistic length of CNTs on the predicted values of $\epsilon_Y$ can be estimated from Eq. (4) as according to,  

\begin{equation}
\Delta\epsilon = - {k_{B}T\over{VK}}\ln( n_{site}/n_{site}^{0}) ,
\end{equation}

where $n_{site}$ is linearly dependent on the length of CNTs. For a $\mu m$
long (10,0) CNT, under the same conditions, the yield or failure strain lowers to about $9\pm 1\%$. Of course, much more total strain energy is needed to yield a longer CNT than the energy needed to yield a shorter CNT. A larger diameter nanotube of the same length may have more activation sites $n_{site}$ for nucleating the defects leading to the yielding of the tube. However, $N$, the number of single processes involved in the yielding, may also increase and offset the effect. This will not be the case if individual single-wall nanotube diameter increases by orders of magnitude, which is highly unlikely.

In Eq.(4), we note that the yield strain of CNTs is linearly dependent on the average kinetic activation energy. The static in-plane activation energy of the first S-W bond rotation, for several armchair and zigzag CNTs, were calculated at zero tension, by rotating the C-C bond in the CNT surface plane and relaxing the system with zero temperature MD simulations. The results are shown in Figure 5. $E_{\nu}$ is found to be around 9.5 eV for large diameter nanotubes. This is about 1 eV different than the activation energy of 8.5 eV computed by Zhang et. al., for a (6,6) nanotube using static tight-binding simulation. \cite{crespi98} This means that the error in the computed averaged activation energy of the multiple kinetic processes leading to the yielding of the nanotube, due to the use of Tersoff-Brenner interactions (as opposed to a different and perhaps more accurate tight-binding or ab-initio interactions) could be estimated to be as high as 1 eV, and can give an estimation of an additional error-bar on our predicted value of the yield strain. According to Eq. (4) 1.0 eV error in the averaged ${\overline E}_{\nu}$ corresponds to about $5\%$ error in the yield strain. The predicted yield strain from the MD simulations using Tersoff-Brenner interactions can further be improved with more accurate tight-binding molecular dynamics simulations in future. Moreover, we note that the static activation energy of the first SW bond rotation is more than twice higher than the averaged kinetic activation energy of successive kinetic S-W bond rotations. This is natural, because the activation energy of the S-W bond rotation, in a strained tube, and in the presence of already existing S-W bond rotations or other defects, is expected to be much lower than the activation energy of the first S-W bond rotation or other defects in an ideal perfect tube. 

Additionally, the plot in Fig. 5 also shows that a small diameter zigzag nanotubes has a smaller activation energy than that of a large diameter zigzag nanotube. No such difference is obvious for the armchair nanotubes. It is expected that the yielding behavior will show a minor diameter dependence for the zigzag nanotubes and no such dependence in the armchair nanotubes. This is shown in the inset in Fig. 5. This also shows that the averaged kinetic activation energy for successive multiple kinetic S-W defect formation processes is a very important parameter in our model and could be improved upon in higher fidelity dynamic or static simulations in future. 

Above analysis shows that the yield strain of a tensile strained CNT is strongly dependent on the temperature and the applied strain rate. Figures 3 and Eq.(4) show that slower strain rates at lower temperatures can be equivalent to faster strain rates but also at higher temperatures. Both the conditions increase the probability of overcoming the activation barriers during the dynamic processes. From Eq.(4), this equivalence can be defined as

\begin{equation}
({\dot {\epsilon}_{1} N \over n_{site} \dot\epsilon_{0}})^{T_1} = ({\dot {\epsilon}_{2} N \over n_{site} \dot\epsilon_{0}})^{T_2} ,
\end{equation}

and the processes can thus be accelerated by elevating the simulation temperatures. In a different context, Voter et. al. have used the high temperature accelerated dynamics to study the diffusion on solid surfaces, \cite{voter00,voter01} which shows that more realistic experimental deposition rates can be reproduced through this scheme. 
Of course the dynamics on CNTs is much simpler compared with other materials, as CNTs are quasi-one dimensional and have simple structures. 

For comparison, the strain rate and temperature dependence of the yielding of CNTs under the compressive stress is also studied using the MD simulations with Tersoff-Brenner potentials. Sideways buckling of the compressed nanotube at lower temperatures, similar to Yakobson et al's MD study at zero temperature,\cite{yakobson96} and a collapsing behavior at higher temperature driven by diamond-like bond formations, as noted by Srivastava et. al. in quantum molecular dynamics simulations were observed.\cite{deepak99} Additionally, S-W type bond rotation defects were also observed in simulations at higher temperatures.\cite{wei01} These defects are found to be irreversible with the removal of the compression induced strain in the system. Our MD simulations show that the diamond-like bond formations and S-W bond rotation defects are strongly correlated with the local buckling instability of the compressed nanotubes, which tend to additionally reduce the barrier heights in the compression case. The complex coupling of the buckling formation with the activated processes, as described in this work, is currently under investigations and will be discussed separately.

Finally, experiments show the tensile strength of 40-50 GPa for SWCNT ropes and MWCNTs, even though their yield strains are different: $5-6 \%$ for SWCNTs and $12 \%$  for MWCNTs. Eq. (4) shows that the yield strain $\epsilon_Y$ is also a function of activation volume $V$ and Young's modulus
$K$. Since $K$ is shown to be about 1 TPa for diverse SWCNTs and MWCNTs, the difference would be related to the activation volume. It is reasonable to expect a smaller activation volume on the outer shell of a MWCNT due to the presence of inner shell CNTs, and the reduced $V$ will increase the value of $\epsilon_Y$ for MWCNTs. These issues are currently under investigations and will be published in future.

Acknowledgment. We thank Dr. M. Baskes for helpful discussions. This work is supported by NASA contract NCC2-5400. DS is supported by NASA contract 704-40-32 to CSC at Ames Research Center.

\begin{figure}
\epsfxsize=5in
\centerline{\epsfbox[0 0 368 434]{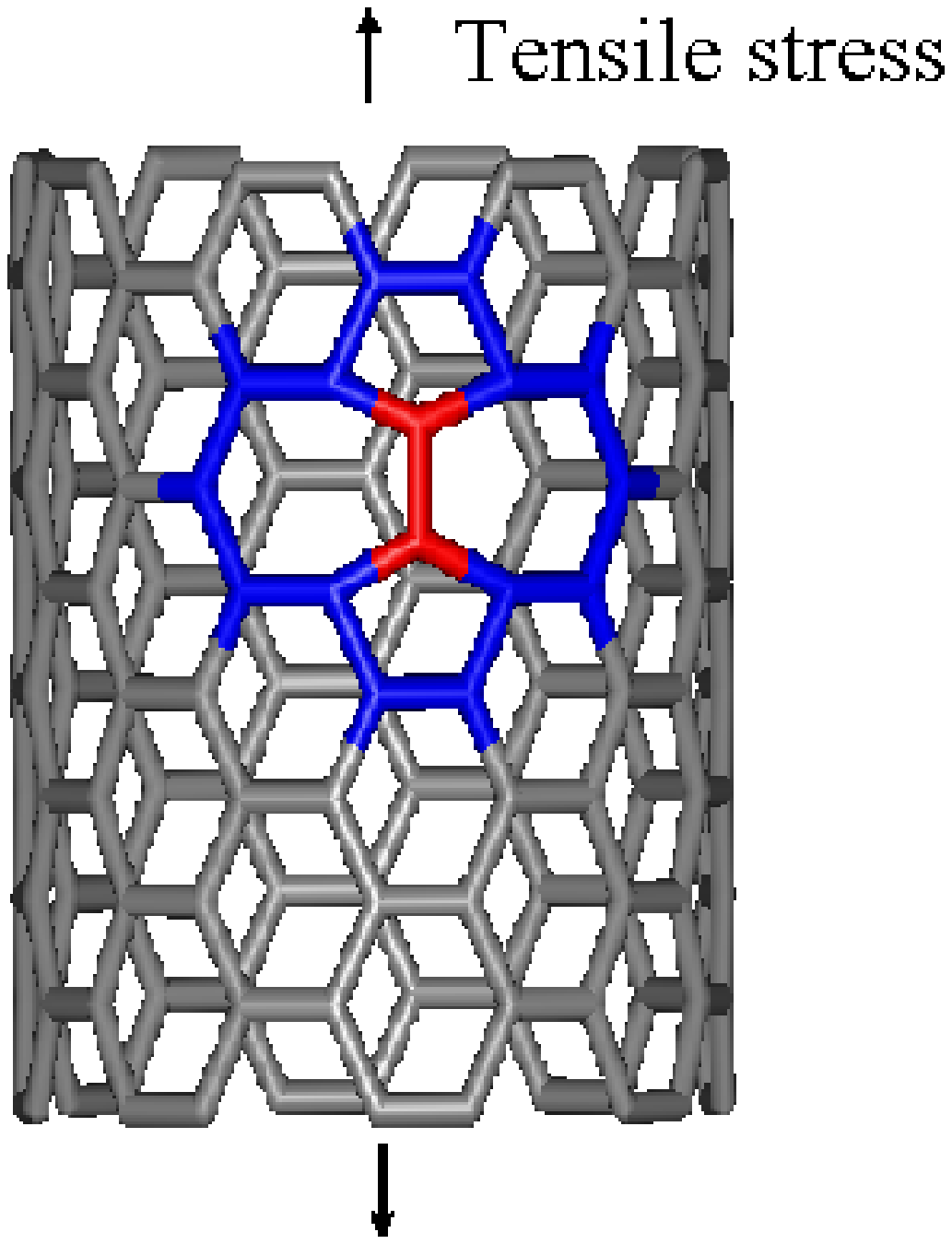}} \caption{A Stone-Wales bond rotation defect shown on an armchair CNT under tension.} \end{figure}

\begin{figure}
\epsfxsize=5.3in
\centerline{\epsfbox[49 40 522 419]{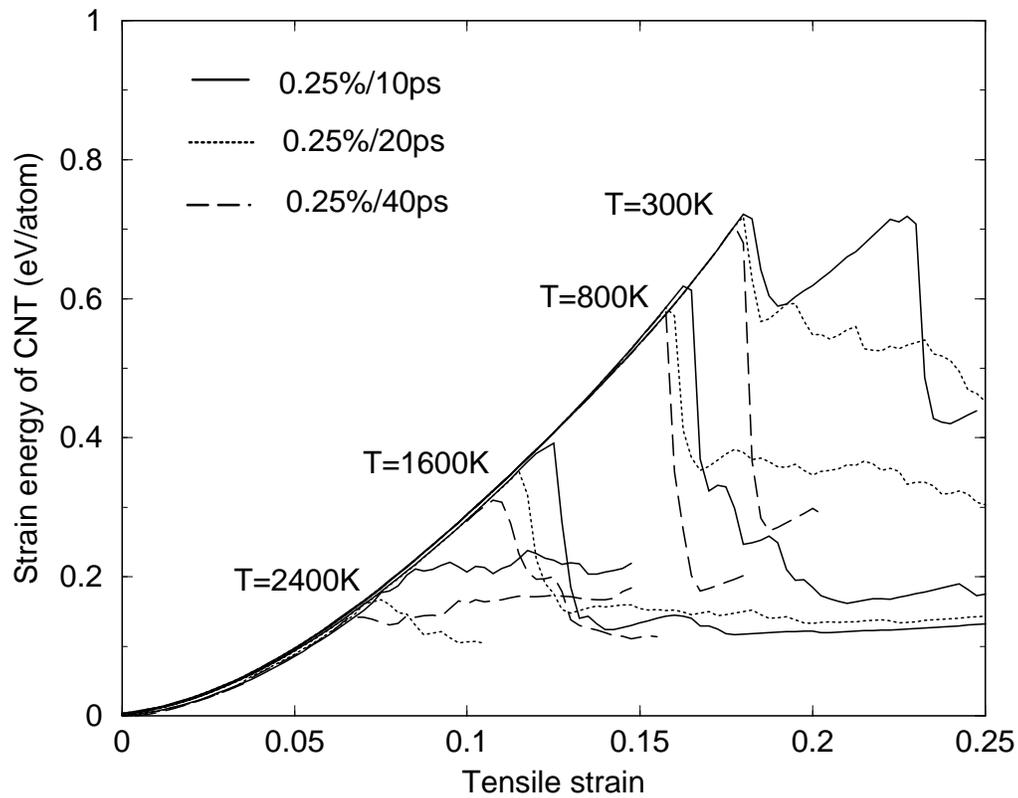}} \caption{The
strain energy per atom as function of tensile strain for CNT(10,0) at
different temperatures and strain rates.} \end{figure}

\begin{figure}
\epsfxsize=5.3in
\centerline{\epsfbox[38 40 521
419]{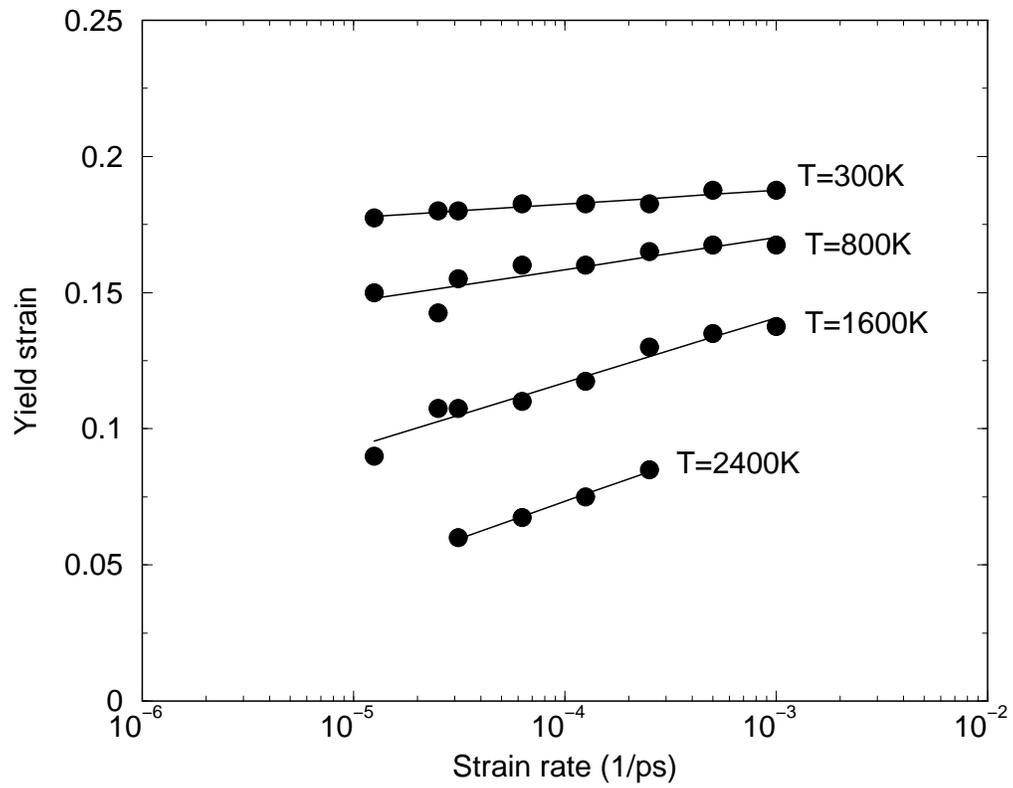}} \caption{Yield strain of (10,0) CNT as a function of strain rate at different temperatures under tensile
stress.} \end{figure}

\begin{figure}
\epsfxsize=5.3in
\centerline{\epsfbox[55 40 524 419]{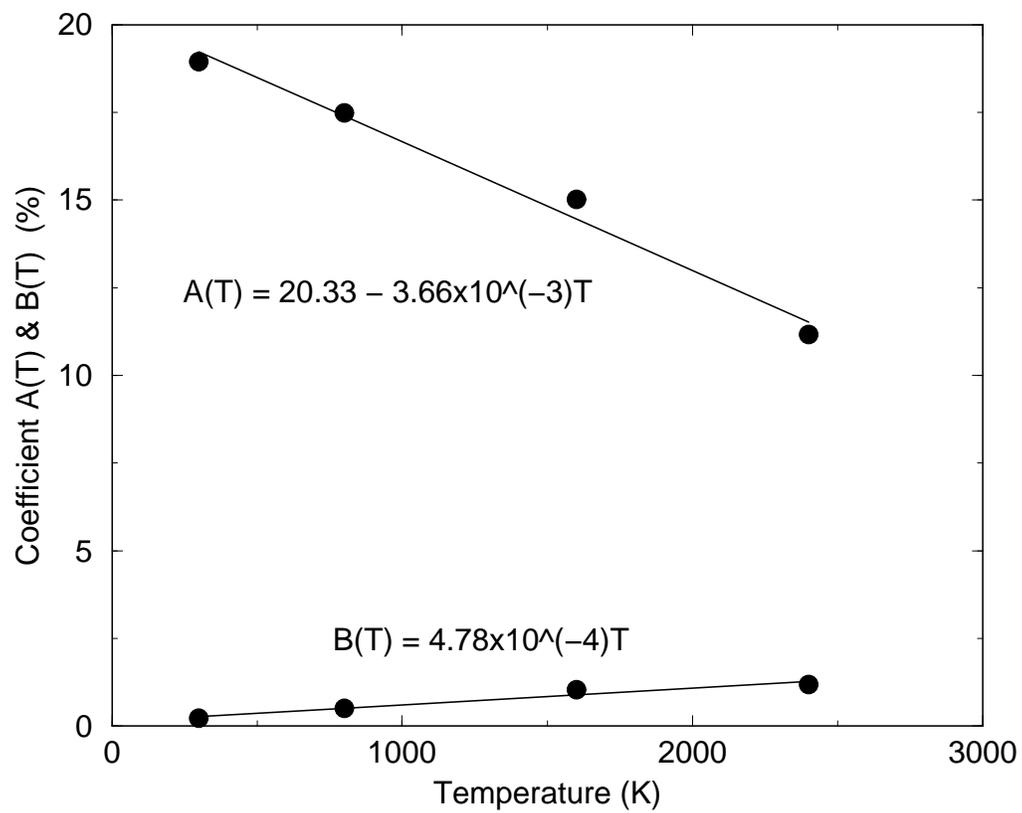}}
\caption{The fitting of A(T) and B(T) with data obtained from Figure
3.} \end{figure}

\begin{figure}
\epsfxsize=5.3in
\centerline{\epsfbox[55 43 511 434]{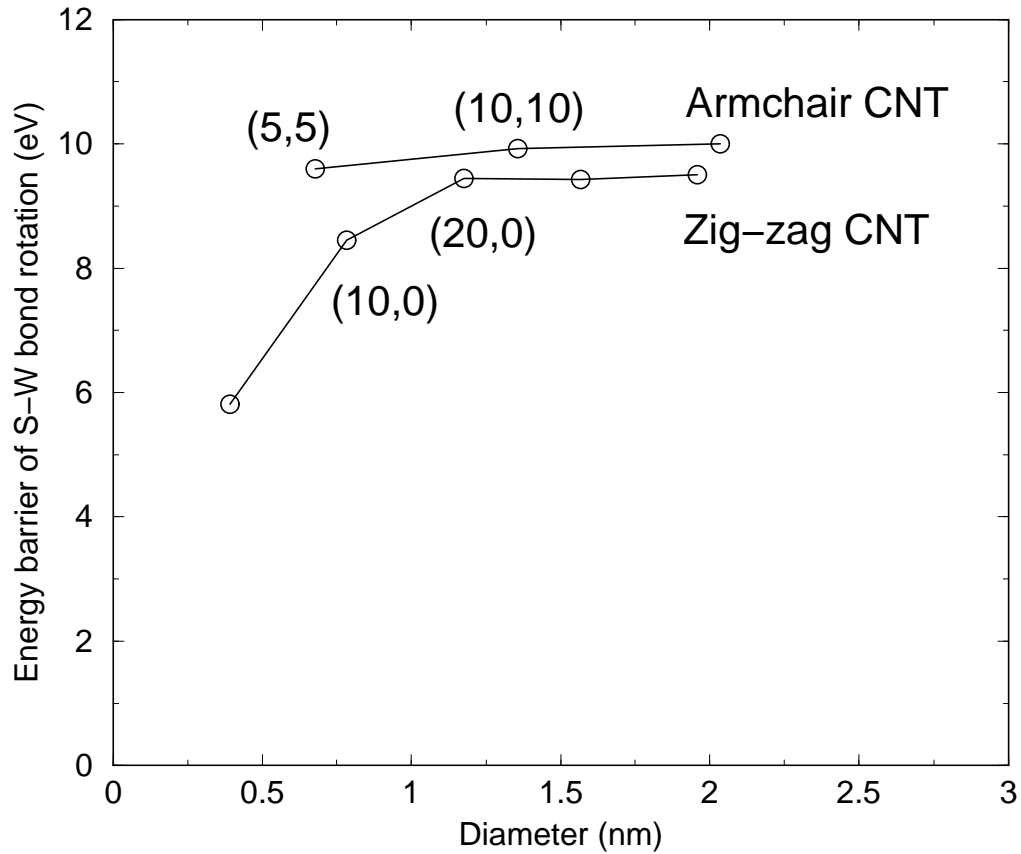}} 
\caption{The static in-plane activation energy of Stone-Wales bond rotations as a function of diameters for zigzag and armchair CNT, using Tersoff-Brenner potentials. Inset (Figure 6 and Figure 7): The yield strain of (10,0), (20,0), (5,5) and (10,10) CNTs as a function of strain rate at T=2400K, from MD simulations.} \end{figure}
\begin{figure}

\epsfxsize=5.3in
\centerline{\epsfbox[49 43 521 434]{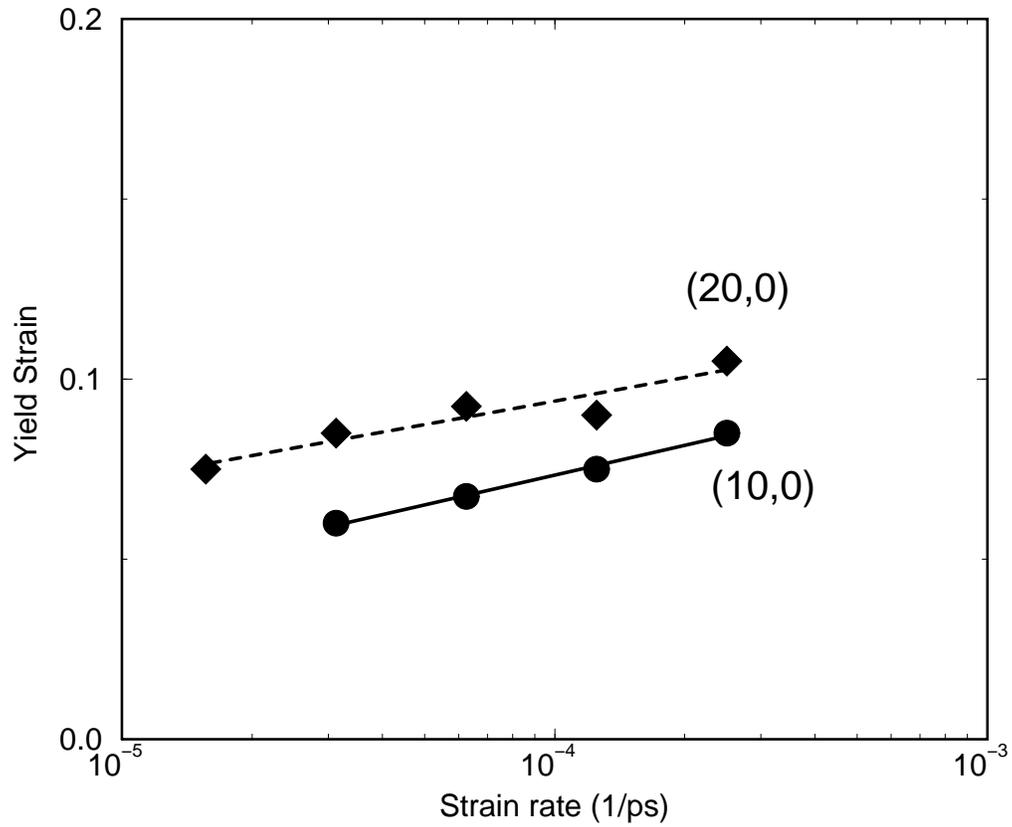}} 
\caption{Inset of Figure 5: The yield strain of (10,0), (20,0) CNTs as a function of strain rate at T=2400K, from MD simulations.} \end{figure}
\begin{figure}
\epsfxsize=5.3in
\centerline{\epsfbox[49 43 521 434]{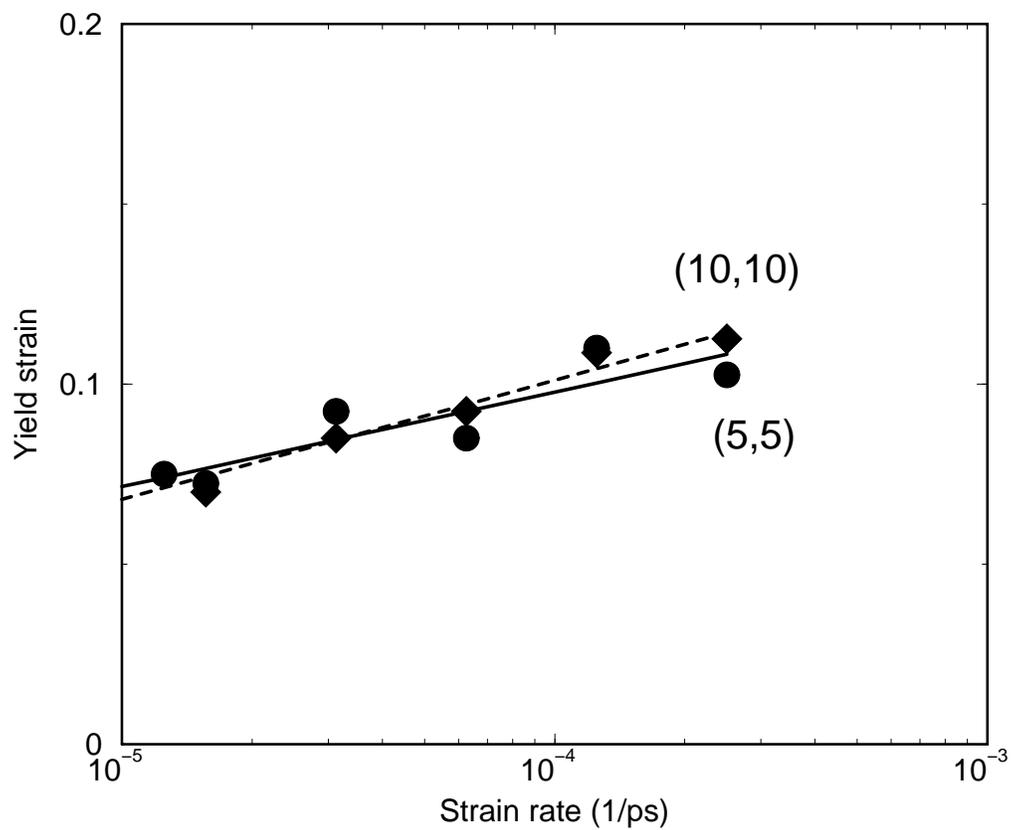}} 
\caption{Inset of Figure 5: The yield strain of (5,5) and (10,10) CNTs as a function of strain rate at T=2400K, from MD simulations.} \end{figure}

\end{document}